\documentclass{ws-ijmpe}
\usepackage[super,compress]{cite}
\usepackage{graphicx}
\usepackage{bm}
\usepackage{amsmath}
\usepackage{amssymb}

\begin{document}

\markboth{K. Murulane, S. Karataglidis, B. G. Giraud}{On the strong local potential limit}

\catchline{}{}{}{}{}
%%%%%%%%%%%%%%%%%%%%%%%%%%%%%%%%%%%%%%%%%%%%%%%%%%%%%%%%%%%%%%%%%%%%

\title{On the strong local potential limit}

\author{K. Murulane}
\address{Department of Physics, University of Johannesburg, P. O. Box 524.\\
Auckland Park, Gauteng, 2006, South Africa\\
murulanek@uj.ac.za}

\author{S. Karataglidis}
\address{Department of Physics, University of Johannesburg, P. O. Box 524.\\
Auckland Park, Gauteng, 2006, South Africa\footnote{Permanent address}\\ and,
School of Physics, University of Melbourne,\\
Victoria, 3010, Australia\\
stevenka@uj.ac.za}

\author{B. G. Giraud}
\address{Institut de Physique Th\'eorique, Centre Etudes, CEA-Saclay, 
Gif-sur-Yvette, 91191, France \\
bertrand.giraud@cea.fr
}

\maketitle

\begin{history}
\received{17 August 2023}
\end{history}

\begin{abstract}
Finite, bound, many-body systems, where the interaction operator,
$V=\sum v_{ij}$, is local and happens to strongly dominate the kinetic energy
operator, $T=\sum t_i$, display a classical limit from diagonalizing $V$ alone,
hence a clear picture of interparticle correlations, such as steric blocking
emerges. This limit exhibits intrinsic symmetries and also
fluctuations from mass formulae. The present work investigates how such
emergent symmetries and fluctuations might be extrapolated to physical
situations where $T$ is reinstated.
\end{abstract}

\keywords{shell structure,nucleon-nucleon interaction}

\ccode{PACS numbers: 13.75.Cs,13.85.-t, 21.60.Cs}

\section{Introduction and basic formalism}
Consider a system of a finite number $A$ of identical constituents with
Hamiltonian, $H=T+V$, where $T=\sum_{i=1}^A p_i^2/(2m)$ is the kinetic energy and
$V=\sum_{i<j} v\left( \left|\mathbf{r}_i-\mathbf{r}_j \right| \right) + v_{cm}$.
Here, $v$ is the local two-body potential, which, for nuclear physics for
instance, is the nucleon-nucleon interaction. (There are many formulations;
see Refs. \refcite{Vo65,Go70,Ma87,Ma89,Wi95} for representative examples.)
The elementary mass is denoted by $m$, and
$\mathbf{p}_i \equiv \left\{ p_{xi},p_{yi},p_{zi}\right\}$ and
$\mathbf{r}_i \equiv \left\{ x_i, y_i, z_i \right\}$ are the momentum and
position of particle $i$, respectively. Finally the term,
$v_{cm}= A m \omega^2 R^2/2$, where 
$\mathbf{R}=A^{-1} \sum_{i=1}^A \mathbf{r}_i \equiv \left\{ X, Y, Z \right\}$,
factorizes the center-of-mass (cm) in a spherical, Gaussian packet at the
origin of the laboratory frame, with energy $\frac{3}{2} \hbar \omega$,
in order to remove translational degeneracy. Without such a term, frequently
omitted in the literature, the ground state of $H$ would not be square
integrable, because of a zero momentum plane wave for its center-of-mass.

Since $T$ is a one-body operator, one usually replaces $V$ by a mean-field
potential or a DFT one\cite{KS}, $U=\sum_{i=1}^A u_i$,
to take advantage of the easy diagonalization of $T+U$.
One then reinstates, as much as possible, correlations brought by $V - U$.
However, in the present work, we exchange the roles of $V$ and $T$, whereupon
we shall show that it makes sense to first diagonalize $V$ and write
$H=V+\lambda T$, where $\lambda$ is a small dimensionless constant
such that $\lambda T$ represents a weak perturbation.
(See Refs. \refcite{Gi82,Gi78,Gi84}, for the virtues of positive (semi-)definite
operators, such as $T$, as perturbations.)

In many cases of physical interest, the physical system under study can be
modeled as driven by a two-body interaction that can be taken as a local,
finite-range, rotationally invariant potential,
$v_{ij}=v\left(  \left|\mathbf{r}_i - \mathbf{r}_j \right| \right)$,
with a short range repulsion, sometimes even a hard core, and a pocket of
attraction at mid-range. The presence or absence of a tail of $v_{ij}$ will
not be important for our arguments in the following. What the diagonalization
of $V$ alone will bring is a clear picture of the role of steric crowding, a
familiar concept in classical physics, but a much less straightforward 
concept for the analysis of full-fledged wave functions, quantum mechanically,
given that sharp positions are washed out by $T$. But we shall show that this
concept of crowding can remain of some utility.

At the ``strong $V$'' limit used herein, it makes no difference whether the
$A$ elements of our system are classical or quantum (boson or fermion) objects.
This stems from the observation that in the limit $\lambda \rightarrow 0$,
the Hamiltonian is driven solely by the potential energy and its spectrum in
this limit is continuous\cite{Gi82}. For the sake of pedagogy and simplicity,
this work ignores spin and other complications such as isospin, Coulomb
effects, and/or mixtures of two or more kinds of particles. This then allows
us to investigate the emergent symmetries of the Hamiltonian
\textit{classically}, and from those evaluations producing a natural
correspondence to quantum mechanical many-body systems. As we shall show, the
purely classical limit provided by $V$ alone illustrates important properties.

The locality of $v$ provides a trivial diagonalization of the full operator
$V$. Indeed, one just needs a search of distinct, \textit{classical} positions,
$\mathbf{s}_i$, and, if one wants to make correspondence with quantum
mechanics, one simply needs a product of very narrow wave packets as a
substitute for the product of $\delta$-functions,
$\prod_{i=1}^A \delta(\mathbf{r}_i - \mathbf{s}_i)$. Those classical positions,
$\mathbf{s}_i$, are the set that minimizes $V$ as a classical function. Note
that, when calculating $\left\langle V \right\rangle$,
exchange terms driven by distinct, very narrow, wave packets are negligible.
(Exchange terms strictly vanish with $\delta$-functions.) Also,
because of $v_{cm}$, any optimal pattern of positions $\mathbf{s}_i$ will
be centered at the origin of the laboratory frame. Rotational degeneracy
remains, but can be handled later, by using a deformed rather than spherical
$v_{cm}$, for example.

The use of classical physics to achieve this may seem counter-intuitive,
but, only important at this stage, is the fact that the diagonalization of
just $V$ generates symmetries and transparently illustrates, often
\textit{through shells}, the steric blocking\cite{Gr12} of particles. This
also gives an intuition of irregularities in the table of binding energies
when the particle number $A$ increases.
(Actually, the use of $\delta$-functions can be alleviated by subtracting
from $V$ a slight perturbation, $\varepsilon\, \left| \phi \rangle \langle \phi \right|$,
where $\varepsilon$ is a small, positive number and $\phi$ is a broad and
flat wave function in $A$ dimensions, then adding that perturbation back
to $\lambda T$\cite{Gi82}. This technicality does not influence physics and
will be understood in the following.)

That $V$ alone can induce shells and also induces mass fluctuations is an
important statement. In the nuclear shell model (for example), the
evidence for shells in the structure of the many-body nucleonic systems comes
from the observation of the magic numbers. A single-particle basis is
constructed to explain the magic numbers. That basis assumes first and
foremost harmonic oscillator single-particle states which are then
supplemented by angular momentum states with spin-orbit splitting, as
well as isospin. Harmonic oscillators are used for the single particle
wave functions for their simplicity and analyticity, making the calculations
tractable. And from that basis, the shell model Hamiltonian may
be diagonalized with respect to many-body configurations within that shell
model basis. The inherent shell structures are an underlying assumption in
that respect, and the potentials derived within the spaces involved are
defined only for those spaces. Woods-Saxon functions would be more
appropriate, but, as the Woods-Saxon potential cannot be solved analytically,
such calculations are highly impractical. Hence, instead of assuming shell
structures \textit{a priori}, can there be a more natural origin of shell
structure?

In Section II, we present a toy model to suggest that
results from $V$ alone are likely to survive the return of $T$. Then, in
Sections III, IV, and V, for the sake of pedagogy, graphical convenience and shorter
computer time, we shows quite a few two-dimensional (2d) results. We then
present a significant number of three-dimensional (3d) results in
Sec. VI, which display an illustrative sampling set of ``$V$ alone''
solutions, and their symmetries and/or irregularities.

Such calculations correspond to a semirealistic situation, with mainly a
trivial Volkov-like potential, made of two Gaussians (2G),
\begin{equation}
v\left( r_{ij} \right)= v_{\text{rep}} \exp\left[ -\mu_{\text{rep}} r^2_{ij} \right] + 
v_{\text{att}} \exp\left[ -\mu_{\text{att}} r^2_{ij} \right],
\label{volkov}
\end{equation}
where $r_{ij} = \left| \mathbf{r}_i - \mathbf{r}_j \right|$ ,
and $v_{\text{rep}}$, $v_{\text{att}}$, $\mu_{\text{rep}}$, and
$\mu_{att}$, specify the strengths and ranges of a
repulsion and an attraction. Such a choice is practical for the calculation
of matrix elements in a harmonic oscillator shell model basis.

Section VII presents our conclusions. And, for what follows, the units are arbitrary in all our results. 

\section{Toy model to study localization}
It is trivial that the ground state of a 1d harmonic oscillator, with
Hamiltonian, $h=p^2/(2\mu)+\frac{1}{2}Kx^2$, lies at energy
$\frac{1}{2} \hbar \sqrt{K/\mu}$
with wave function $\pi^{-\frac{1}{4}} \beta^{-\frac{1}{2}} e^{-x^2/(2\beta^2)}$,
where the width $\beta$ relates to $\mu$ through $\beta^2=\hbar/\sqrt{K\mu}$. We shall
investigate whether, when $\mu$ is finite and large, delocalization effects
remain small, of order $1/\sqrt{\mu}$.

\begin{figure}
\centerline{\scalebox{0.50}{\includegraphics*{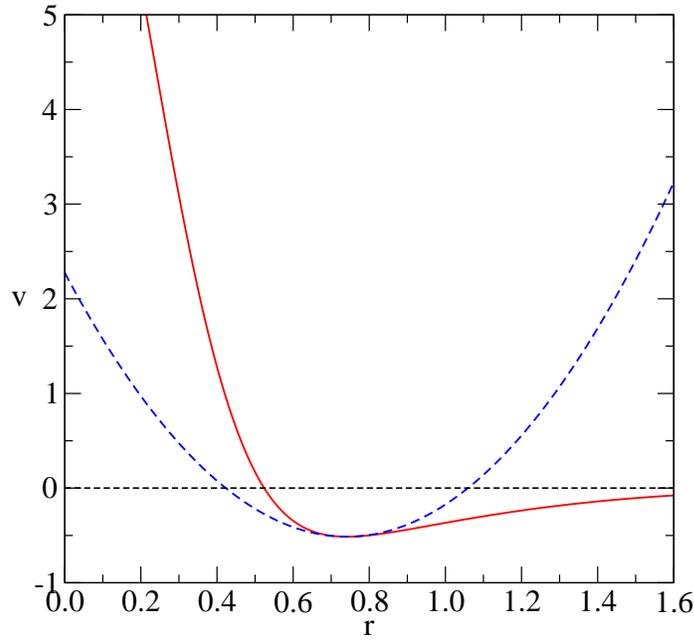}}}
\caption{\label{potandappxm} The toy potentials used to compare the Gaussian approximations with
the full solutions. The toy potential around its minimum is portrayed by the solid line while
the dashed line portrays its harmonic approximation.}
\end{figure}

\begin{figure}
\centerline{\scalebox{0.50}{\includegraphics*{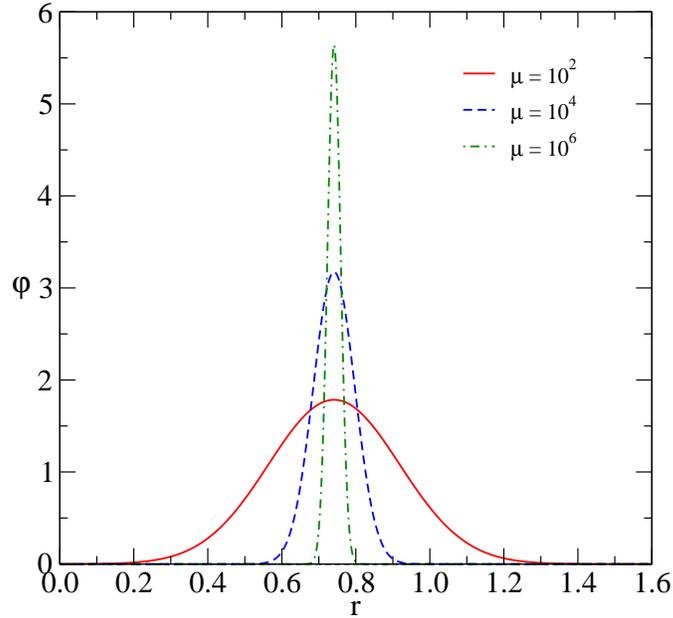}}}
\caption{\label{GaussShrink} Narrowing Gaussians as the mass increases.
The three cases shown are $\mu=10^2$ (solid line), $\mu=10^4$ (dashed line), and $\mu=10^6$ (dot-dashed
line).}
\end{figure}

We set $\hbar=1$, and, in a transparent notation, the equation for relative
motion reads,
\begin{align}
  - \frac{1}{2\mu}\frac{d^2\psi\left( r \right)}{dr^2} + \left( 9 e^{-9r^2} -
  e^{-r^2} \right) \psi\left( r \right) & = \eta_{rel} \psi\left( r \right),
  \nonumber \\
\psi\left( 0 \right) & = 0.
\label{rltv}
\end{align}
(Remember that the cm factorises out into a Gaussian.) We are only
interested in this relative motion equation, Eq.~(\ref{rltv}),
with its relative coordinate $r=\left|\mathbf{r}_2 -\mathbf{r}_1 \right|$, relative momentum
$p_{rel}=\left| \mathbf{p}_2-\mathbf{p}_1 \right|/2$ and relative mass $\mu=m/2$.
An expansion of the potential from its minimum, $v_0=-0.51320$ at
$r_0=0.741152$, yields the harmonic approximation of Eq.(\ref{rltv}),
\begin{equation}
-\frac{1}{2\mu}\frac{d^2\varphi\left(r \right)}{dr^2} + \left[ v_0 + \frac{1}{2}
  \left( \frac{d^2v_0}{dr^2} \right) \left( r - r_0 \right)^2 \right]
\varphi\left( r \right) = \eta_h \, \varphi\left( r \right),
\end{equation}
where $d^2v_0/dr^2 =10.1485$ is the second derivative of the potential at its
minimum. The potential with its harmonic approximation is shown in 
Fig.~\ref{potandappxm}. It will be instructive to compare the radial wave $\psi$ and
the Gaussian $\varphi$, although the latter does not strictly vanish when
$r=0$. But both narrow in width when $\mu$ increases.
We show in Fig.~\ref{GaussShrink} the Gaussians obtained when
$\mu=10^2,10^4$, and $10^6$. The narrowing of the Gaussians with increasing $\mu$ is
clearly shown.

\begin{figure}
\centerline{\scalebox{0.50}{\includegraphics*{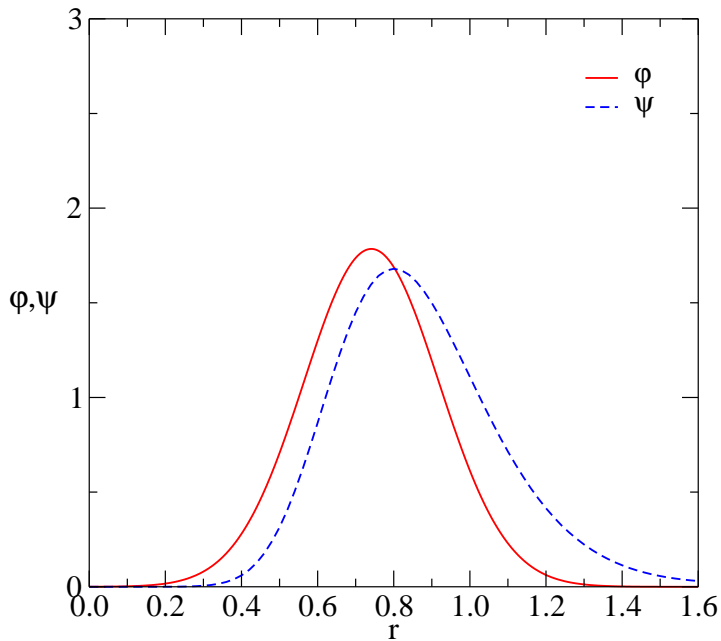}}}
\caption{\label{compare1} The case for $\mu=10^2$: the wave packet, 
the relative wave function, is denoted by the dashed line; the solid line
portrays its Gaussian equivalent. The same potentials as in 
Fig.~\ref{potandappxm} are used.}
\end{figure}

We show in Fig.~\ref{compare1} how close from each other the true 
relative function, at energy $\eta_{rel} = - 0.369560$, and its Gaussian
approximation, at energy $\eta_h = - 0.353916$, are when $\mu=10^2$. Note
that the former wave packet is slightly shifted towards greater $r$ compared
to the latter. This occurs because the $v$ potential there is much less 
confining than the parabolic branch.

\begin{figure}
\centerline{\scalebox{0.50}{\includegraphics*{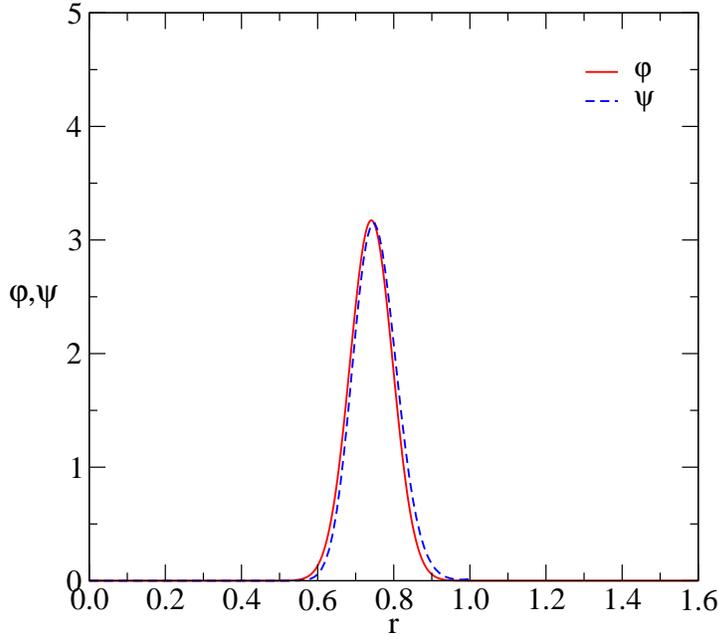}}}
\caption{\label{compare2} The case for $\mu=10^4$: 
the relative wave function is given by the dashed line, while the solid line portrays its Gaussian partner. The same 
potentials as in Fig.~\ref{potandappxm} are used.}
\end{figure}

The case where $\mu=10^4$ is shown in Fig.~\ref{compare2}, where the true wave
packet, at energy $-0.497446$, shrinks like the harmonic one, at energy
$-0.497272$, and a tiny shift remains. But then, when $\mu=10^6$, the wave
packets are so close to each other that the shift becomes negligible (and so not shown).
The localization process is at work and is driven by the scale order,
$\mu^{-1/2}$.

\section{Toy 2G potential for two-dimensional systems}
\begin{figure}
\centerline{\scalebox{0.45}{\includegraphics*{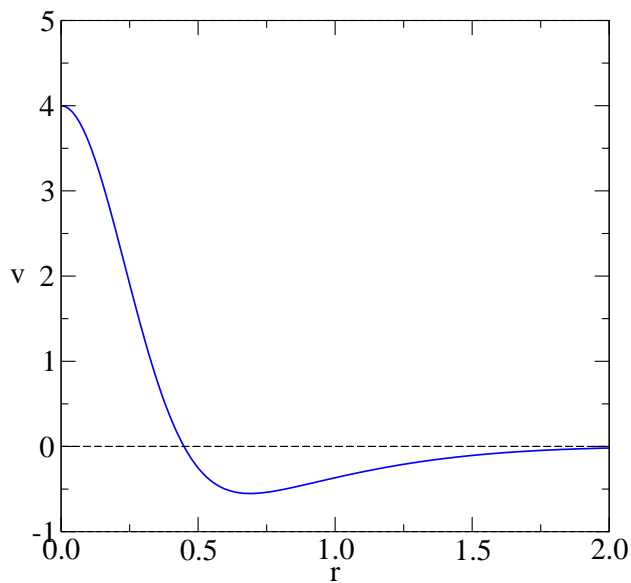}}}
\caption{\label{forcGLJ} The toy 2G potential used in the 2d calculations.}
\end{figure}

In the following, we first use the schematic, scalar potential,
\begin{equation}
  v_{ij} = 5 \exp\left[-9 \left( \mathbf{r}_i -\mathbf{r}_j \right)^2 \right] -
  \exp \left[ - \left( \mathbf{r}_i - 
\mathbf{r}_j \right) ^2 \right],
\label{2G_2d_pot}
\end{equation}
a difference of two Gaussians (henceforth denoted as ``2G''). This is
displayed by the solid line in Fig.~\ref{forcGLJ}.
The bottom of the 2G potential lies at $E_2=-0.5523$ with interparticle distance
$r_2=0.6898$. 

Such interparticle distances and bindings also give the correct results
for the links of equilateral triangles, the trivial solution when $A=3$. 

For the results to follow, we begin the calculation with a random
distribution of the particles in a defined area (2d calculations) or volume
(3d calculations later in this paper) and the two-body interaction is
applied to any two particles in the system. With successive two-body
interactions applied to the system, we seek the minimal energy solution,
the resulting configuration of which we term the optimal configuration.
Finally, the coordinates are translated such that the cm is at the origin
of the coordinate system.

\section{Shell structure}
\begin{figure}
\centerline{\scalebox{0.45}{\includegraphics*{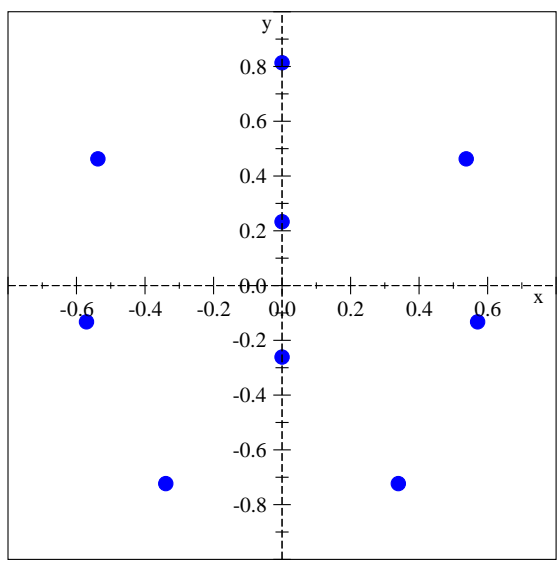}}}
\caption{\label{solu2G9} Optimal configuration obtained with the 2G potential
for $A=9$. The energy is $-13.318$.}
\end{figure}

\begin{figure}
\centerline{\scalebox{0.45}{\includegraphics*{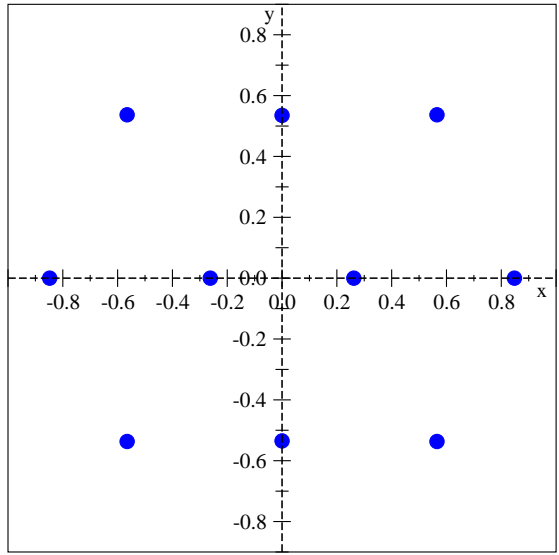}}}
\caption{\label{solu2G10} As for Fig.~\ref{solu2G9}, but for $A=10$.
The energy is $-15.807$.}
\end{figure}

\begin{figure}
\centerline{\scalebox{0.45}{\includegraphics*{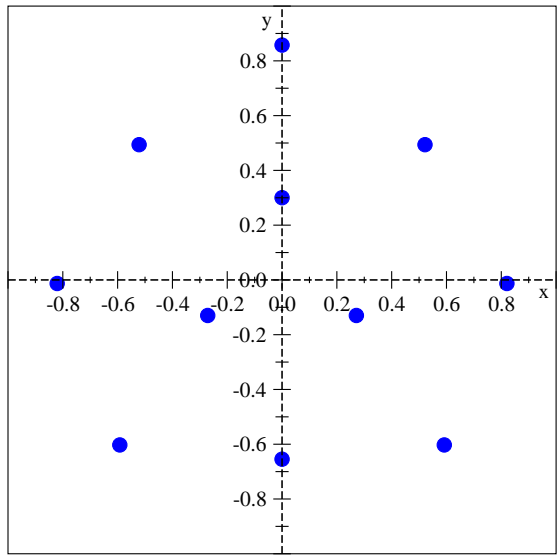}}}
\caption{\label{solu2G11} As for Fig.~\ref{solu2G9}, but for $A=11$.
The energy is $-18.620$.}
\end{figure}

\begin{figure}
\centerline{\scalebox{0.45}{\includegraphics*{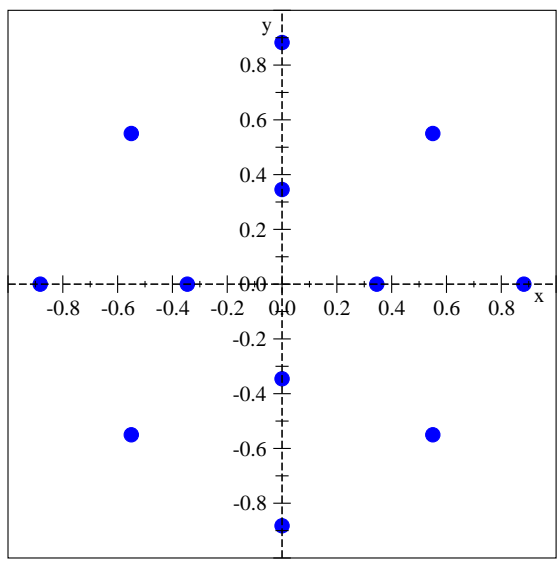}}}
\caption{\label{solu2G12} As for Fig.~\ref{solu2G10}, but for $A=12$.
  The energy is $-21.457$.}
\end{figure}

We begin with a nine-particle system, under the influence of the 2G
potential, through to the twelve-particle system. That sequence is displayed
in Figs.~\ref{solu2G9} to \ref{solu2G12}, and exhibits an increase in
particle number in either the inner shell or the outer shell. This is not
entirely predictable: the evolution of the configurations to the minimal
energy solution is classical. The Pauli Principle
is not imposed on the system and so there is no maximum set number of
particles in each shell as dictated by angular momentum quantum numbers in
quantal systems. For this sequence,
$9 \leqslant A \leqslant 12$, the numbers in the inner shell are
$2, 2\left( 3 \right), 3, 4$ while those in the outer shell
are $7, 8\left( 7 \right), 8, 8$. The corresponding energies are
$-13.318, -15.807\left( -15.774 \right), -18.620, -21.457$. The
anomalous case, $A = 10$, is a result of a competing solution,
corresponding to a very slightly excited configuration (as
given in parentheses) compared to the minimal energy configuration. That
excited configuration has 3 particles in the inner shell and 7 in 
the outer.

The existence of an alternative configuration for $A = 10$ corresponding
to the excited energy is not unique. It also
occurs for $A = 12$, whereby the excited energy configuration (for which
the energy is $-21.337$) has 3 particles in the inner shell 
instead of 4. With that configuration there is also a breaking of symmetry;
the configuration morphs from having two symmetry axes to one. 

There is no doubt that the emergence of shell structure is observed at
this point, although that differs from the usual behaviour where the inner
shell is essentially
stable and with growth occurring in the outer shell. One may speculate that
the introduction of Coulomb effects may enhance the population of outer
shells. The presence in Figs.~\ref{solu2G9} through \ref{solu2G12}, of quite
a few
quasi-alignments of particles and some parallel alignments is also
noteworthy. Further, one may
also conjecture that several more or less equilateral triangles are emerging.
This suggests that
some hexagonal crystallization may be at work.

\section{Shells of many-particle systems}

\subsection{From two to three shells}

\begin{figure}
\centerline{\scalebox{0.45}{\includegraphics*{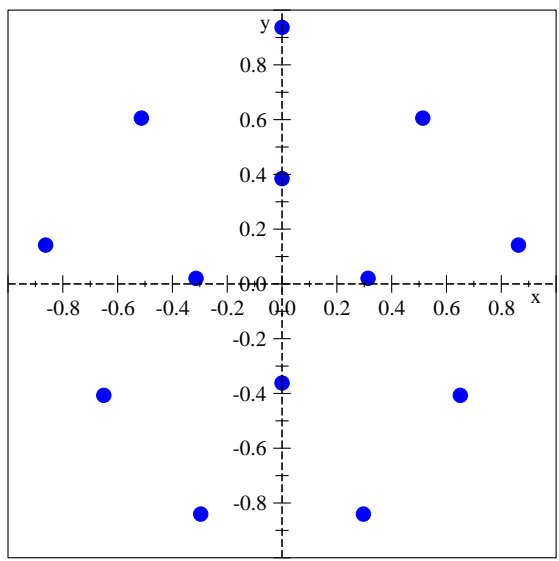}}}
\caption{\label{solu2G13} Optimal configuration using the 2G interaction
  for $A=13$. The energy is $-24.347$.}
\end{figure}

\begin{figure}
\centerline{\scalebox{0.45}{\includegraphics*{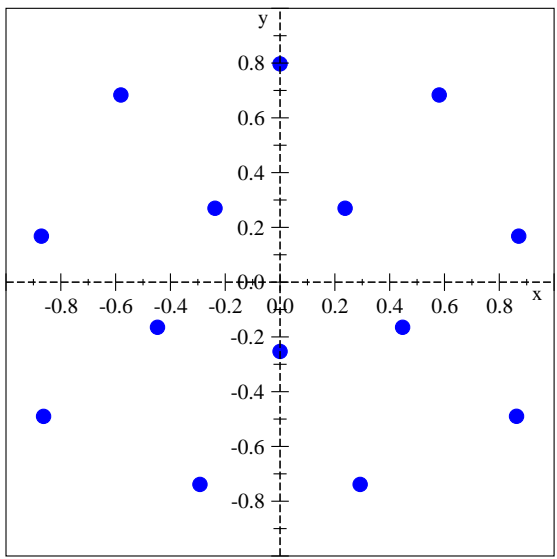}}}
\caption{\label{solu2G14} As for Fig. \ref{solu2G13}, but for $A=14$.
  The energy is $-27.298$.}
\end{figure}

We begin with the optimal configurations for $A=13, 14$,  being the minimal
energy solutions for those particle numbers, as obtained using the 2G potential.
Those solutions are shown in Figs.~\ref{solu2G13} and \ref{solu2G14},
respectively.

\begin{figure}
\centerline{\scalebox{0.45}{\includegraphics*{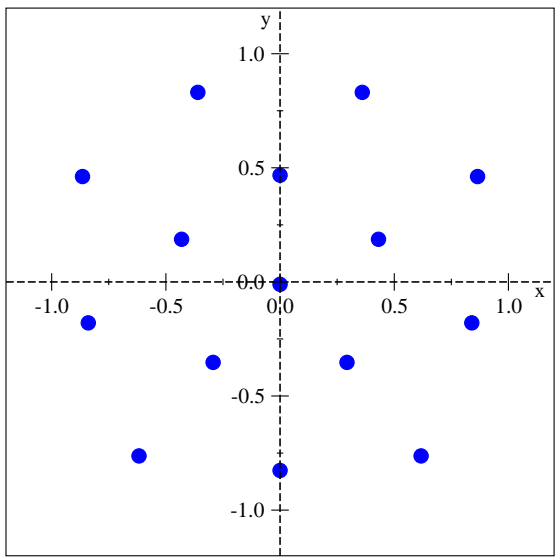}}}
\caption{\label{solu2G15} As for Fig. \ref{solu2G13}, but for $A = 15$.
  The energy is $-30.378$.}
\end{figure}

With the addition of one particle, giving the $A=15$ system, one observes
the emergence of a third shell, manifesting in the
first instance as a particle at the centre-of-mass. This is shown in
Fig.~\ref{solu2G15}.
This is observed also in the optimal configurations for
$16 \leqslant A \leqslant 19$.

\begin{figure} 
\centerline{\scalebox{0.45}{\includegraphics*{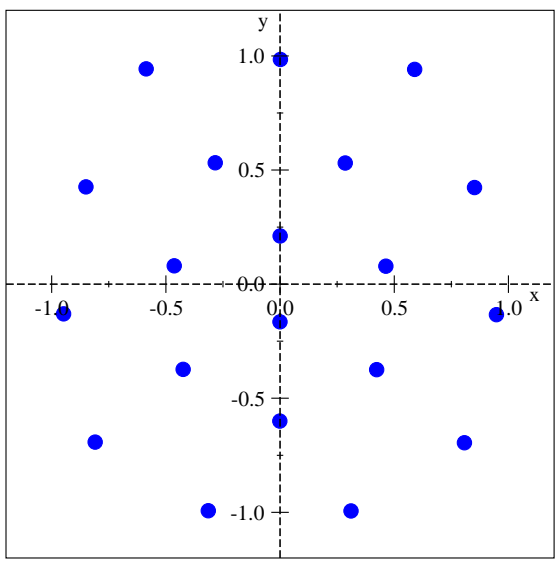}}}
\caption{\label{solu2G20} As for Fig. \ref{solu2G13}, but for $A=20$.
  The energy is $-47.628$.}
\end{figure}

The situation changes for $A=20$, as shown in Fig.~\ref{solu2G20}. The innermost
shell gains an additional particle, which
results in there being no particle in the centre-of-mass, but two particles
nearby, defining an axis of symmetry.
\begin{figure}
\centerline{\scalebox{0.45}{\includegraphics*{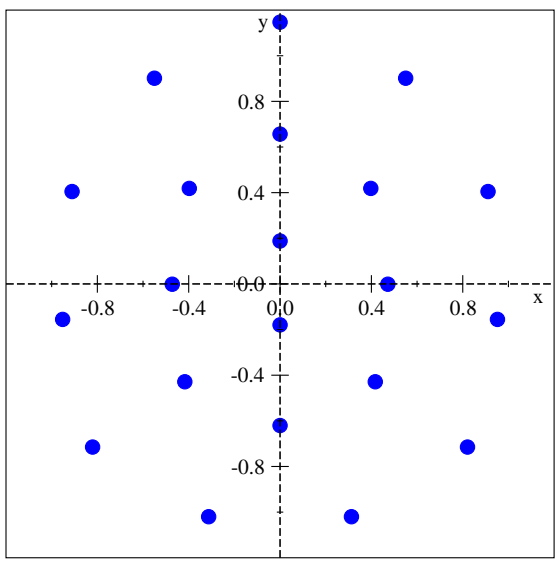}}}
\caption{\label{solu2G21} As for Fig.~\ref{solu2G13}, but for $A=21$.
  The energy is $-51.331$.}
\end{figure}
From $A=20$ to $A=21$, the latter being shown in Fig~\ref{solu2G21}, one
observes that the addition of the particle occurs in the middle shell. 

\begin{figure}
\centerline{\scalebox{0.45}{\includegraphics*{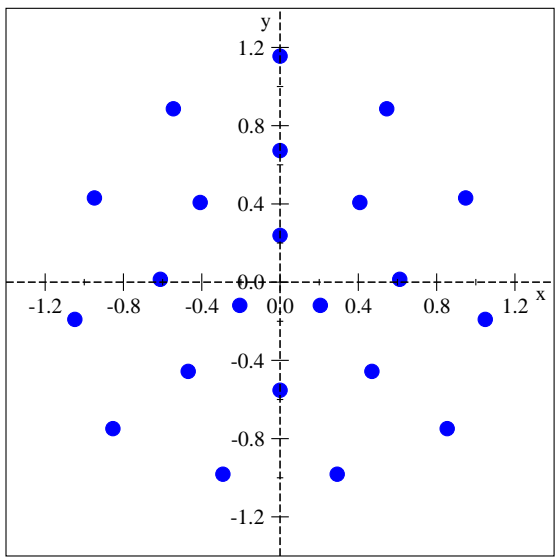}}}
\caption{\label{solu2G22} As for Fig. \ref{solu2G13}, but for $A=22$.
  The energy is $-55.083$.}
\end{figure}

\begin{figure}
\centerline{\scalebox{0.45}{\includegraphics*{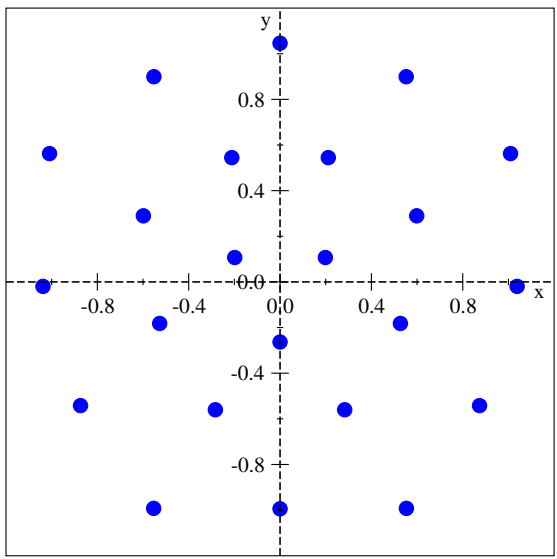}}}
\caption{\label{solu2G23} As for Fig. \ref{solu2G13}, but for $A=23$.
  The energy is $-58.993$.}
\end{figure}

The addition of one more particle to form the $A=22$ system, shown in
Fig.~\ref{solu2G22}, results in a change in the innermost shell, forming
a triangle, while the other two shells remain largely undisturbed.
The particle added to the $A=22$ system, forming the $A=23$ system as
shown in Fig.~\ref{solu2G23}, places that additional particle in the outer
shell.
It can be noted that in all cases, there exists at least one axis of symmetry.

\subsection{From three to four shells and beyond}

\begin{figure}
\centerline{\scalebox{0.45}{\includegraphics*{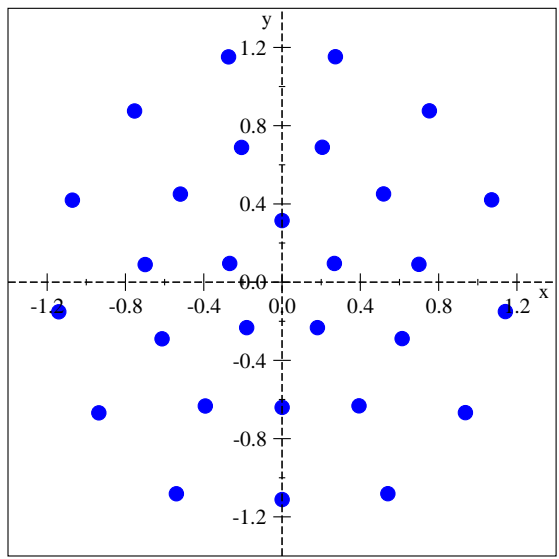}}}
\caption{\label{solu2G29} As for Fig. \ref{solu2G13}, but for $A=29$.
  The energy is $-83.933$.}
\end{figure}

\begin{figure}
\centerline{\scalebox{0.45}{\includegraphics*{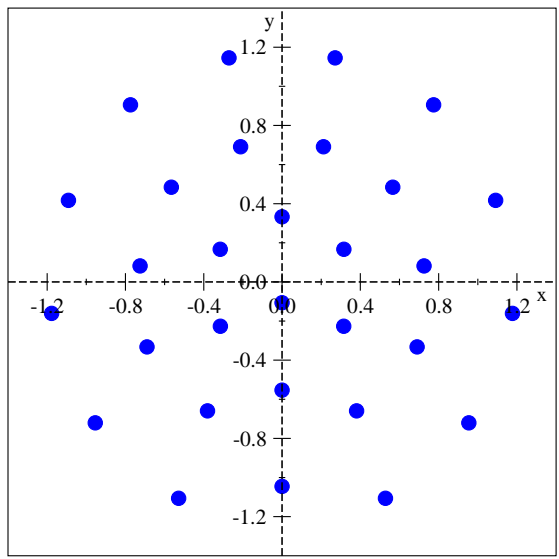}}}
\caption{\label{solu2G30} As for Fig. \ref{solu2G13}, but for $A=30$.
  The energy is $-88.357$.}
\end{figure}

The emergence of the fourth shell occurs for $A=30$. Figs.~\ref{solu2G29}
and \ref{solu2G30} show this transition,
where the $A=29$ system is the largest which exhibits three shells, with a
pentagonal innermost shell, and $A=30$ showing
the fourth shell, manifest as a single particle near the centre.
The population of each shell for $A=30$ is $1,6,10,13$, proceeding from the
innermost to the outermost. However,
the shells are not truly distinct: there is a little ambiguity in the
assigning of the number in the second shell, as one of
the particles may be in the third shell. Yet that particle, on the $y$ axis,
defines the shell as complete, in terms of
a circular symmetry of the second shell consistent with the other shells,
and so we ascribe it to the second shell. Further
work is required as there is a lack of convexity in the outermost shell, and
a fully convex solution may exist.

\begin{figure}
\centerline{\scalebox{0.45}{\includegraphics*{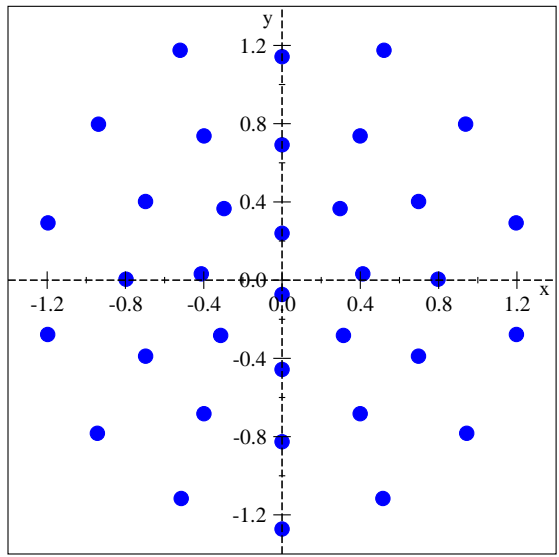}}}
\caption{\label{solu2G35} As for Fig. \ref{solu2G13}, but for $A=35$.
  The energy is $-111.773$.}
\end{figure}

Figure~\ref{solu2G35} shows the optimal configuration of the system of
35 particles, with energy $-111.773$. This is the largest system which
still has one particle, at or near the centre-of-mass, defining the
innermost shell.
The shell populations are 1, 8, 12, and 14, from the innermost to the
outermost shells, respectively, with the same
ambiguity in the inner shells as displayed in Fig.~\ref{solu2G30}. While
one may interpret the configuration as
having two particles in the innermost shell and seven in the next shell,
the definition of the outermost shell requiring
the same average interparticle distance between all particles in that shell
would favour the populations as stated with 
one particle only in the innermost shell.

\begin{figure}
\centerline{\scalebox{0.45}{\includegraphics*{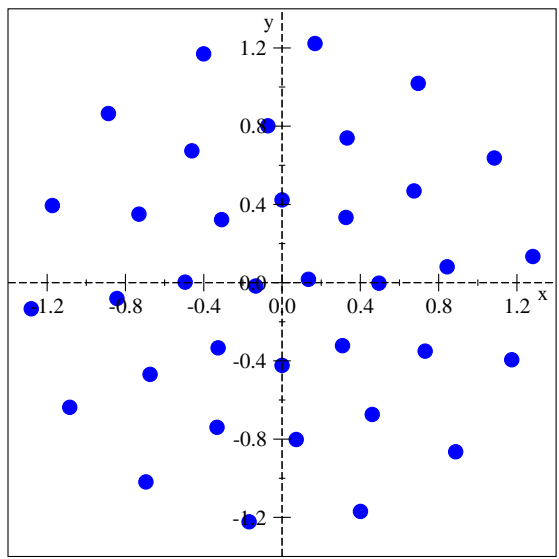}}}
\caption{\label{solu2G36} As for Fig. \ref{solu2G13}, but for $A=36$.
  The energy is $-116.701$.}
\end{figure}

It is in the next system, $A = 36$, with energy $-116.701$, where
two particles in the innermost shell become distinct. That is shown in 
Fig.~\ref{solu2G36}.
The shell populations are 2, 8, 12 and 14. In this case, there are four
distinct shells, but no real axis of symmetry. 
The particles near the $x$ axis suggest the possibility of a symmetry axis
but that is not well-defined. Notice, however that the centre-of-mass turns
out to be a symmetry centre of the whole pattern.

\begin{figure}
\centerline{\scalebox{0.45}{\includegraphics*{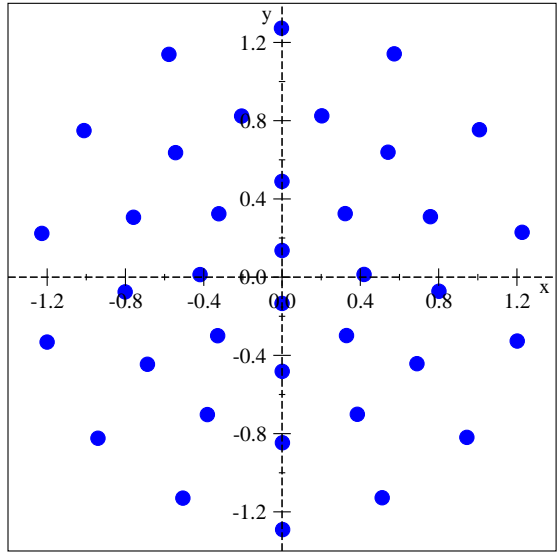}}}
\caption{\label{solu2G37} As for Fig. \ref{solu2G13}, but for $A=37$.
  The energy is $-121.706$.}
\end{figure}

For $A=37$, for which the optimal configuration with energy $-121.706$ is
shown in Fig.~\ref{solu2G37}, a pair of particles in the innermost shell is
still observed.
The populations from the innermost to the outermost shells
are 2, 8, 13, and 14. In this case, a symmetry axis is recovered.

While not displayed, we have also obtained
results for $A=40$ and 41, which show more population of the innermost shells,
with populations of 3 and 4, respectively, while retaining four shells overall.

We have also obtained preliminary results for the transition from four
to five shells, with the transition
occurring at $A=48$, showing 6 particles in the innermost shell. That is also
the case for
$A=49$. At $A=50$, the single
particle at the centre returns, with 7 particles in the next shell. This
requires further investigation.

Generally, one observes the growth of shells beginning with the innermost
shell until saturation occurs,
typically with five or six particles, after which the new shell emerges with
a single particle at or close to the centre. The
rest of the shells grow to accommodate the behaviour in the centre. This is
intuitive: one expects that the
particles influenced by the greatest number of interactions to be near the
centre rather than the outer shells, which
redistribute the populations to account accordingly. This may lead to a
classical correspondence of magic numbers
which, in quantal many-body systems, are attributed exclusively to angular
momentum. It should be noted that these configurations display $SO(2)$ symmetry,
indicating angular momentum emerges as a natural symmetry. (The analysis
will be the subject of a forthcoming paper.)

\section{From two to three dimensions}

It is clear from the results for the optimum configurations in two dimensions
that the
potential [Eq.~(\ref{2G_2d_pot})] gives rise to concentric shells centred at
the centre-of-mass. We
now extend that to the case of three dimensions, using the potential, based
on Eq.~(\ref{volkov}),
\begin{equation}
  v_{ij} = 9 \exp\left[ -9\left( \mathbf{r}_i - \mathbf{r}_j \right)^2 \right] -
  \exp\left[ - \left(
\mathbf{r}_i - \mathbf{r}_j \right)^2 \right],
\label{2g3}
\end{equation}
which we denote as ``2G3'', taking into account the change in steric
crowding. We also, for these cases, give histograms for
the distributions of the radii from the centre-of-mass of each particle to
assist in illustration.
As before, we begin with a random distribution of particles in a given
volume, then apply the two-body 2G3 interaction to the system and allow the system to converge to the
optimal, minimum energy, configuration.

\begin{figure}
\centerline{\scalebox{0.7}{\includegraphics*{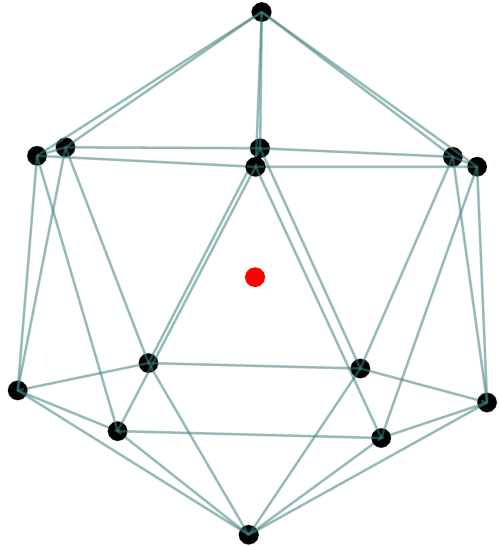}}}
\caption{\label{fig3d_15} Optimal configuration for a system of 15 particles
  in 3 dimensions.
  The energy is $E = -38.0945$. The different colours and lines for the
  shells serve as a guide only: red denotes the 
inner shell while black denotes the outer shell.}
\end{figure}

\begin{figure}
\centerline{\scalebox{0.65}{\includegraphics*{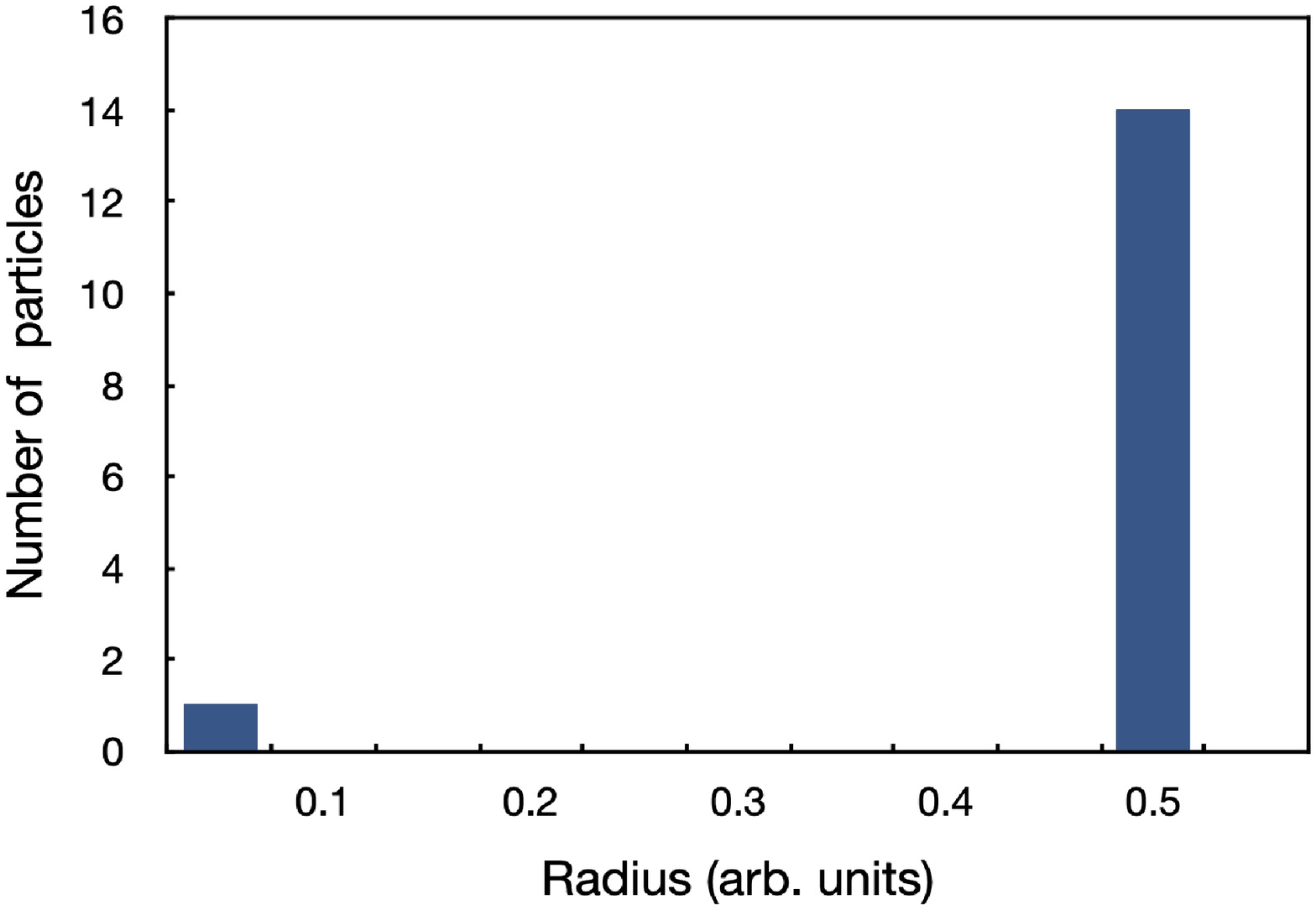}}}
\caption{\label{hist_15} Histogram of particle radii for the $A=15$ system.}
\end{figure}

We begin with the $A = 15$ particle system by way of illustration.
The optimal configuration obtained is
shown in Fig.~\ref{fig3d_15}, for which the energy is $E = -38.0945$.
The histogram of the radii is shown in Fig.~\ref{hist_15}. Therein, the
configuration is 
a particle at the centre surrounded by a shell of radius 0.5.

\begin{figure}
\centerline{\scalebox{0.65}{\includegraphics*{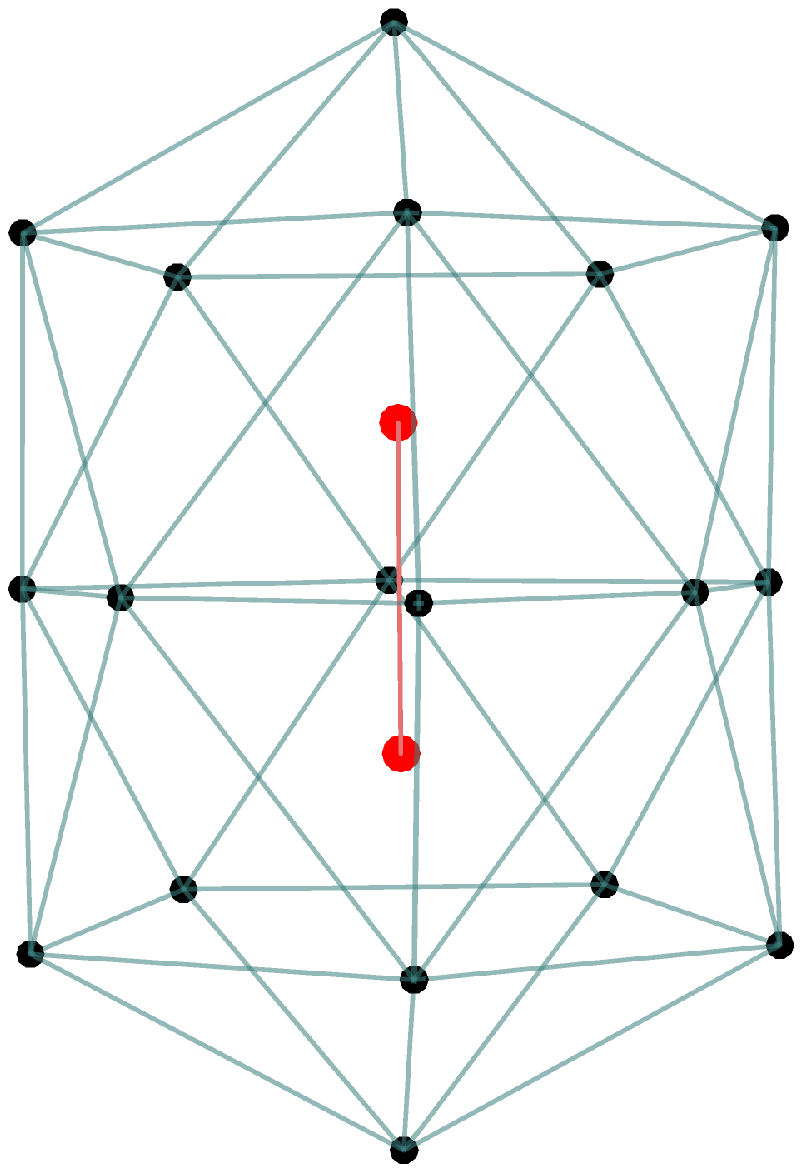}}}
\caption{\label{fig3d_20} As for Fig.~\ref{fig3d_15}, but for $A = 20$.
  The energy is $E = -61.4145$.}
\end{figure}

\begin{figure}
\centerline{\scalebox{0.65}{\includegraphics*{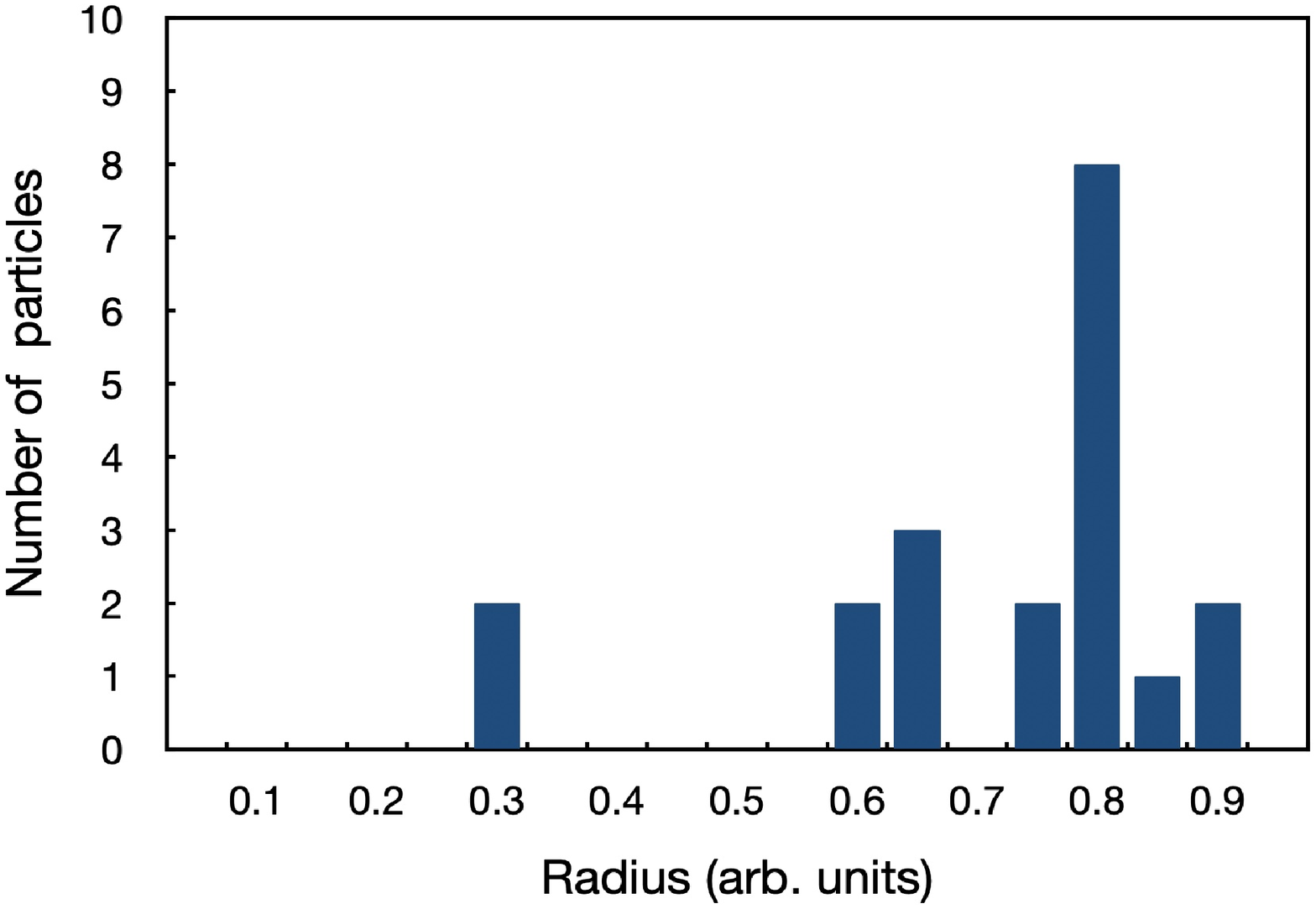}}}
\caption{\label{hist_20} Histogram of particle radii for the $A=20$ system.}
\end{figure}

The optimal configuration for the $A=20$ system is shown in Fig.~\ref{fig3d_20},
for which the corresponding histogram for the particle radii is given in
Fig.~\ref{hist_20}.
For the $A=20$ system, the inner shell, defined by two particles,
clearly defines an axis
of symmetry, and the particles in the outer shell are distributed
evenly along that axis of symmetry,
defining more a cylindrical shell rather than a spherical one. That is
reflected in the wider distribution
of radii of the particles in the outer shell as shown in Fig.~\ref{hist_20}.

\begin{figure}
\centerline{\scalebox{0.65}{\includegraphics*{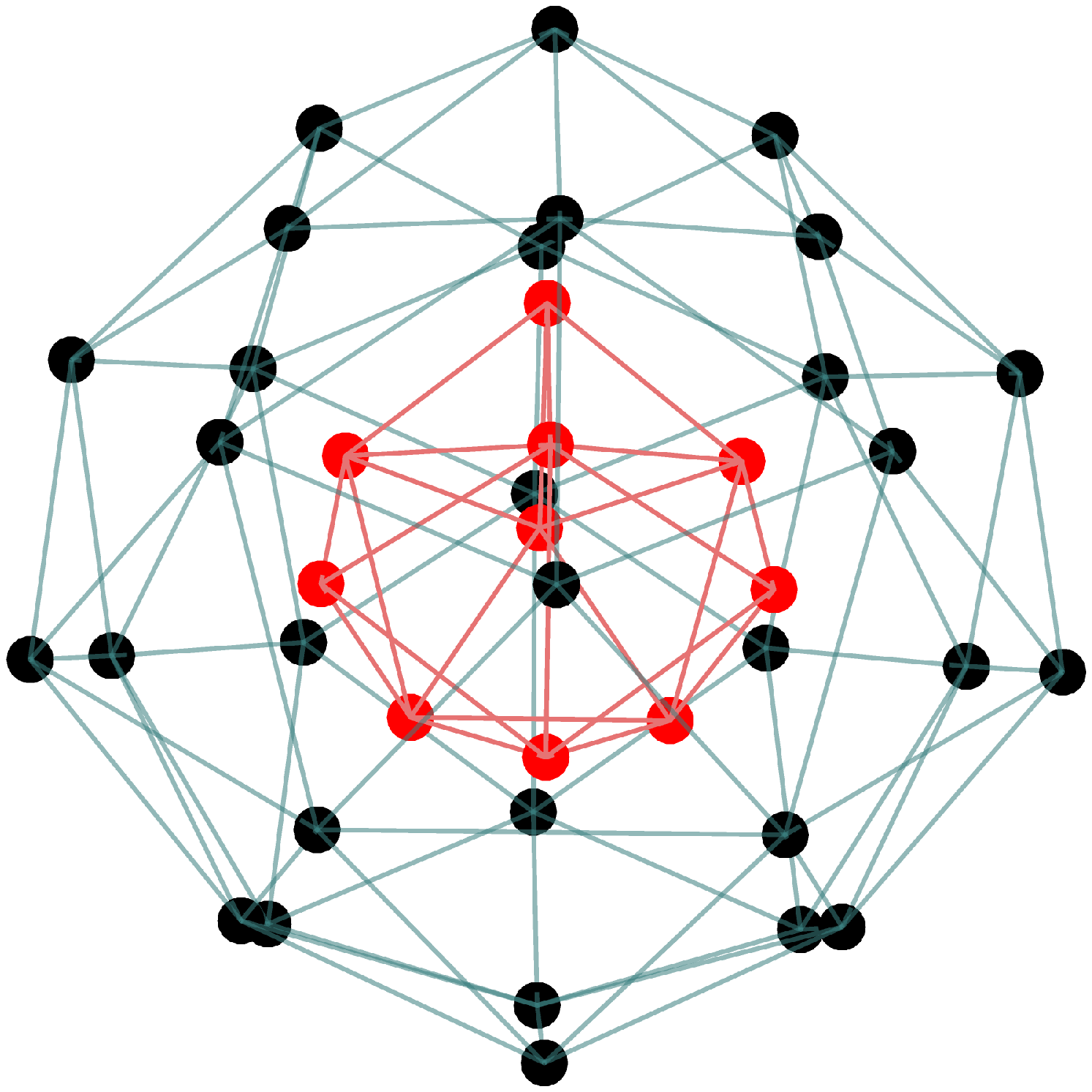}}}
\caption{\label{fig3d_40} As for Fig.~\ref{fig3d_15}, but for a 40 particle
  system. The energy is $E = -193.6239$.}
\end{figure}

\begin{figure}
\centerline{\scalebox{0.65}{\includegraphics*{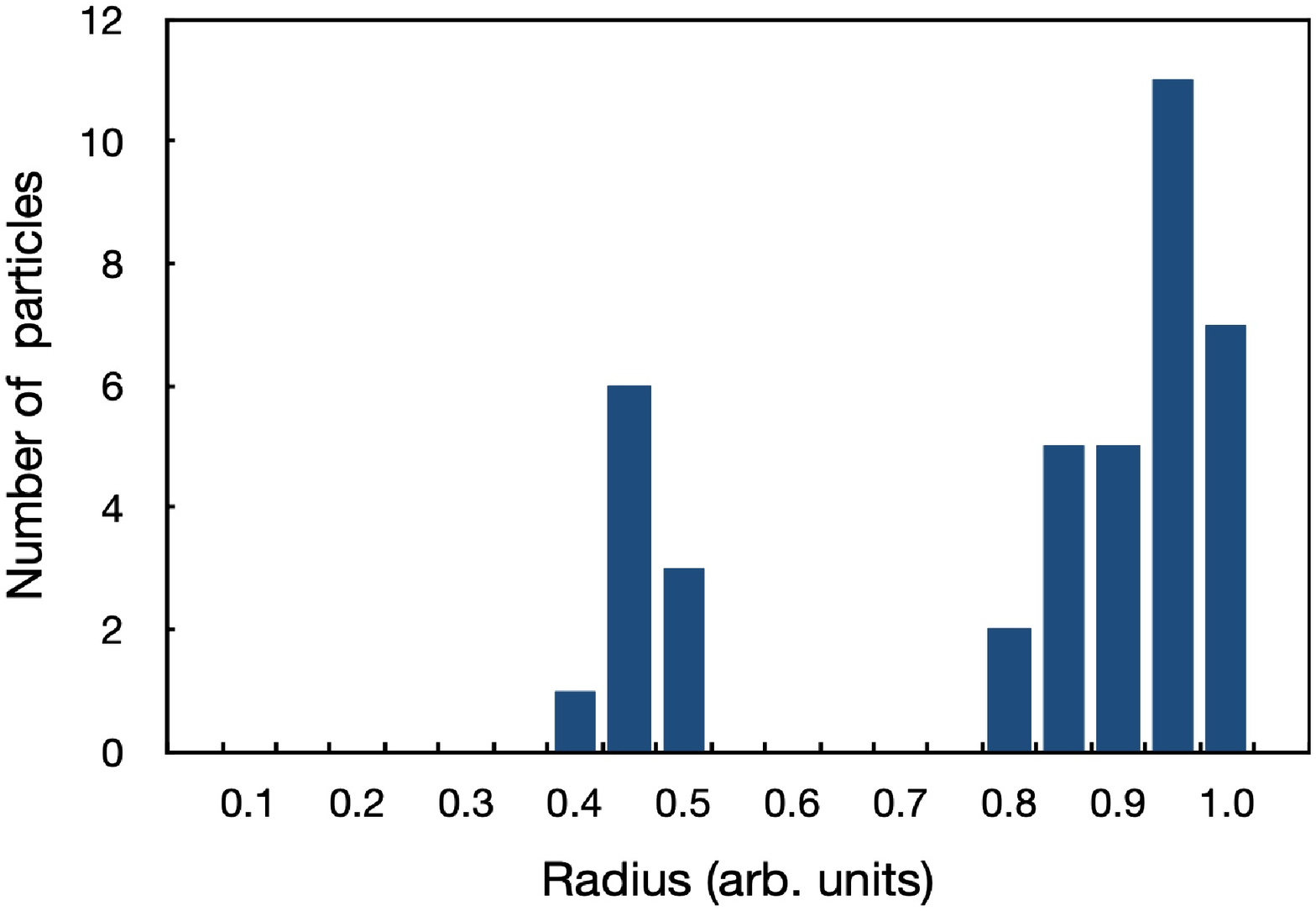}}}
\caption{\label{hist_40} Histogram of particle radii for the $A=40$ system.}
\end{figure}

Fig.~\ref{fig3d_40} displays the optimal configuration for $A=40$,
with the histogram of
the particles' radii shown in Fig.~\ref{hist_40}. The energy is
$E = -192.6239$.
Again, there are two distinct concentric spherical shells in this
configuration, with the radius of
the outer shell roughly twice the radius of the inner one, as shown
in the histogram.

\begin{figure}
\centerline{\scalebox{0.65}{\includegraphics*{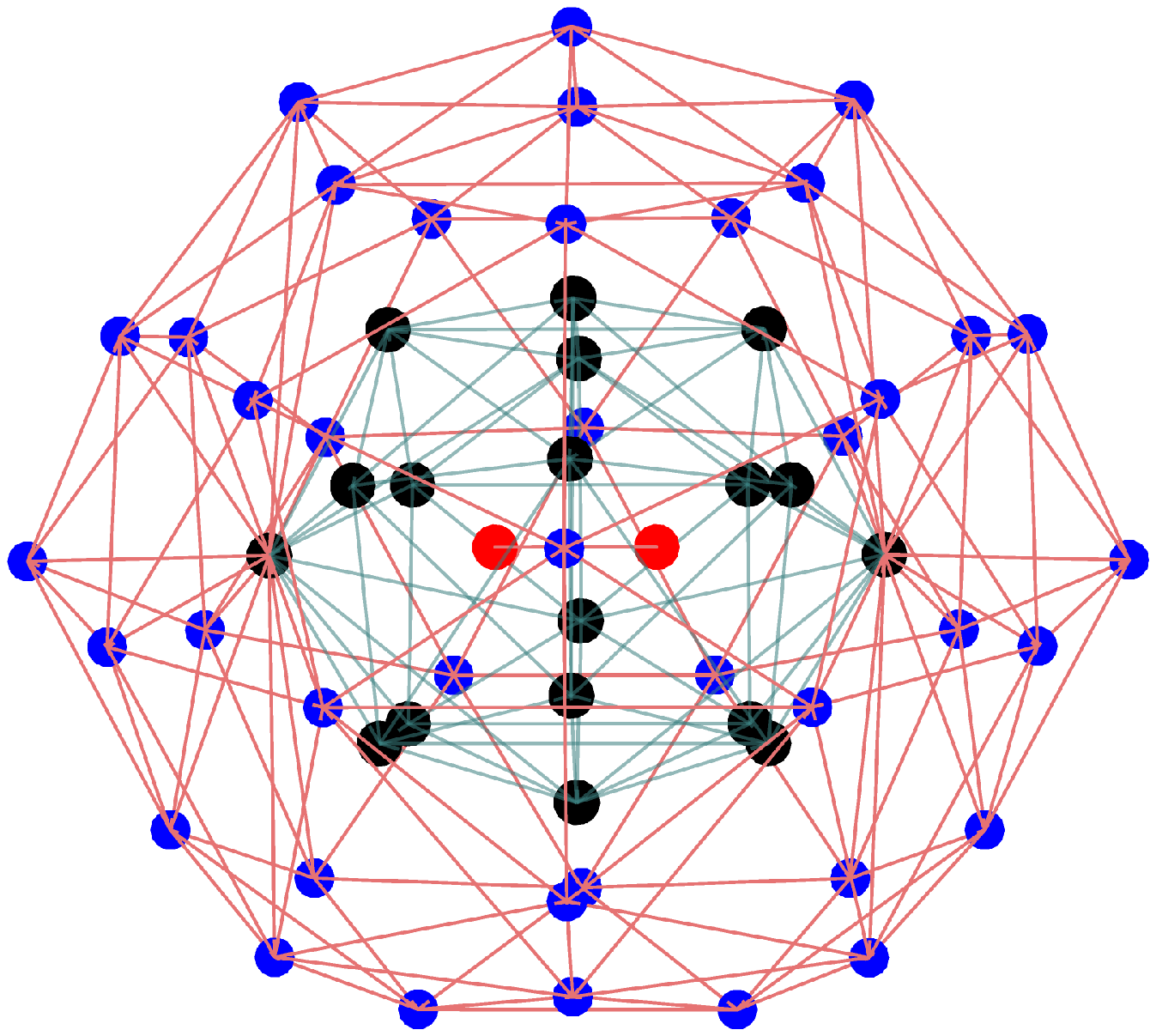}}}
\caption{\label{fig3d_60} As for Fig.~\ref{fig3d_15}, but for $A=60$. The
  energy is $E = -374.8513$. The red and black shells
  denote the innermost and middle shells, respectively, while the outer shell
  is in blue.}
\end{figure}

\begin{figure}
\centerline{\scalebox{0.65}{\includegraphics*{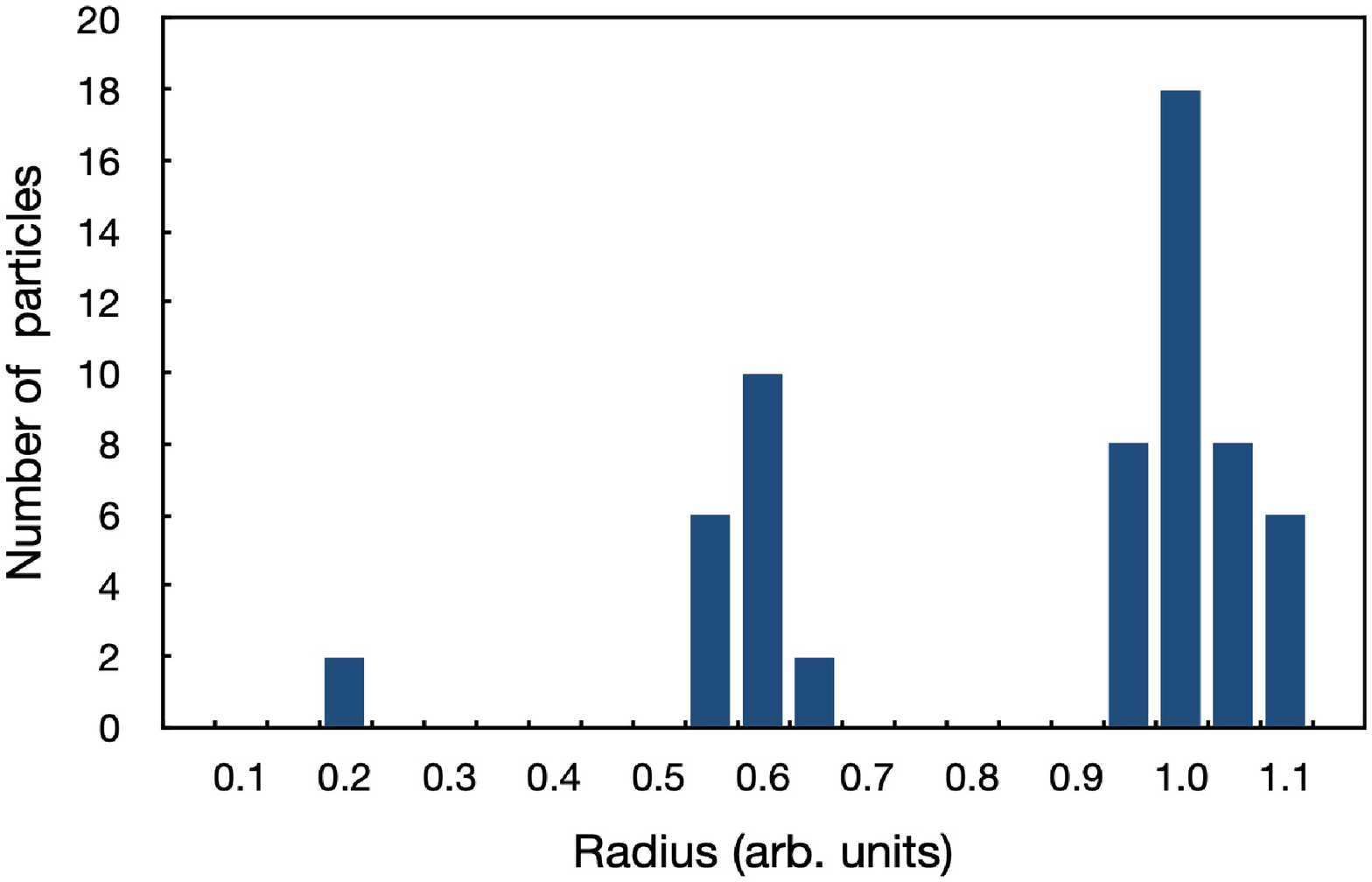}}}
\caption{\label{hist_60} Histogram of particle radii for the $A=60$ system.
  As indicated by the spacing of the
particles' radii, this system has a configuration of three shells.}
\end{figure}

The optimal configuration and histogram of the particles' radii for the
$A=60$ system are shown
in Figs.~\ref{fig3d_60} and \ref{hist_60}, respectively.
For $A=60$ there are now three distinct shells, as confirmed by the radii.
The innermost shell has only two particles but
with the much larger number of particles compared to $A=20$, the cylindrical
symmetry is
relaxed and the shells are now spherical. 

\begin{figure}
\centerline{\scalebox{0.85}{\includegraphics*{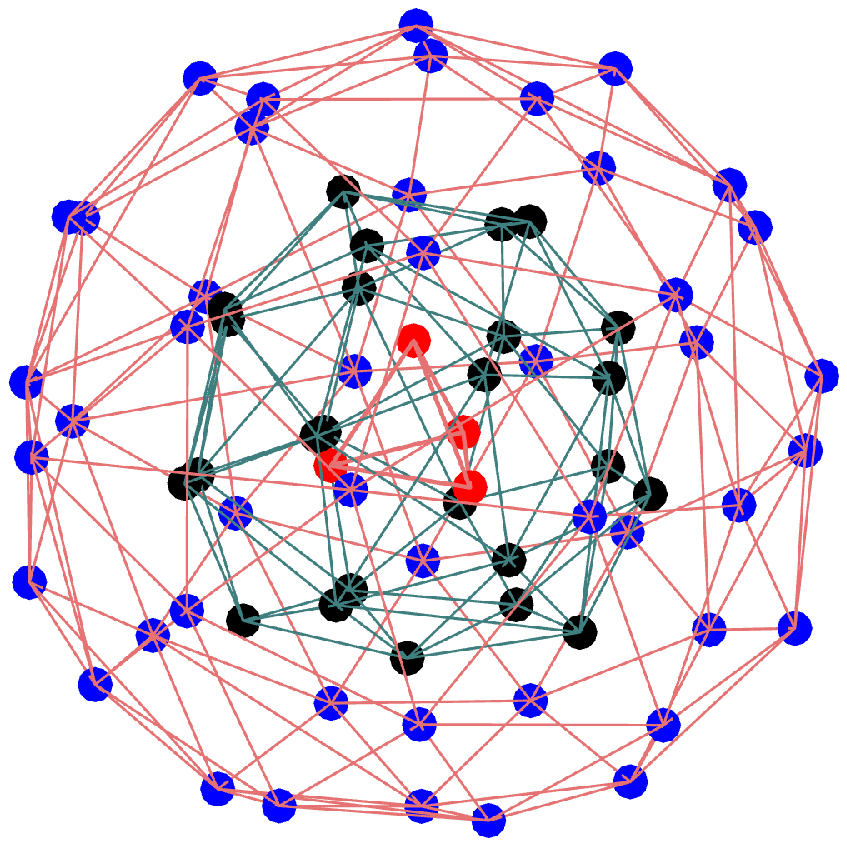}}}
\caption{\label{fig3d_75} As for Fig.~\ref{fig3d_60}, but for the $A=75$
  system. The energy is $E = -539.6275$.}
\end{figure}

\begin{figure}
\centerline{\scalebox{0.65}{\includegraphics*{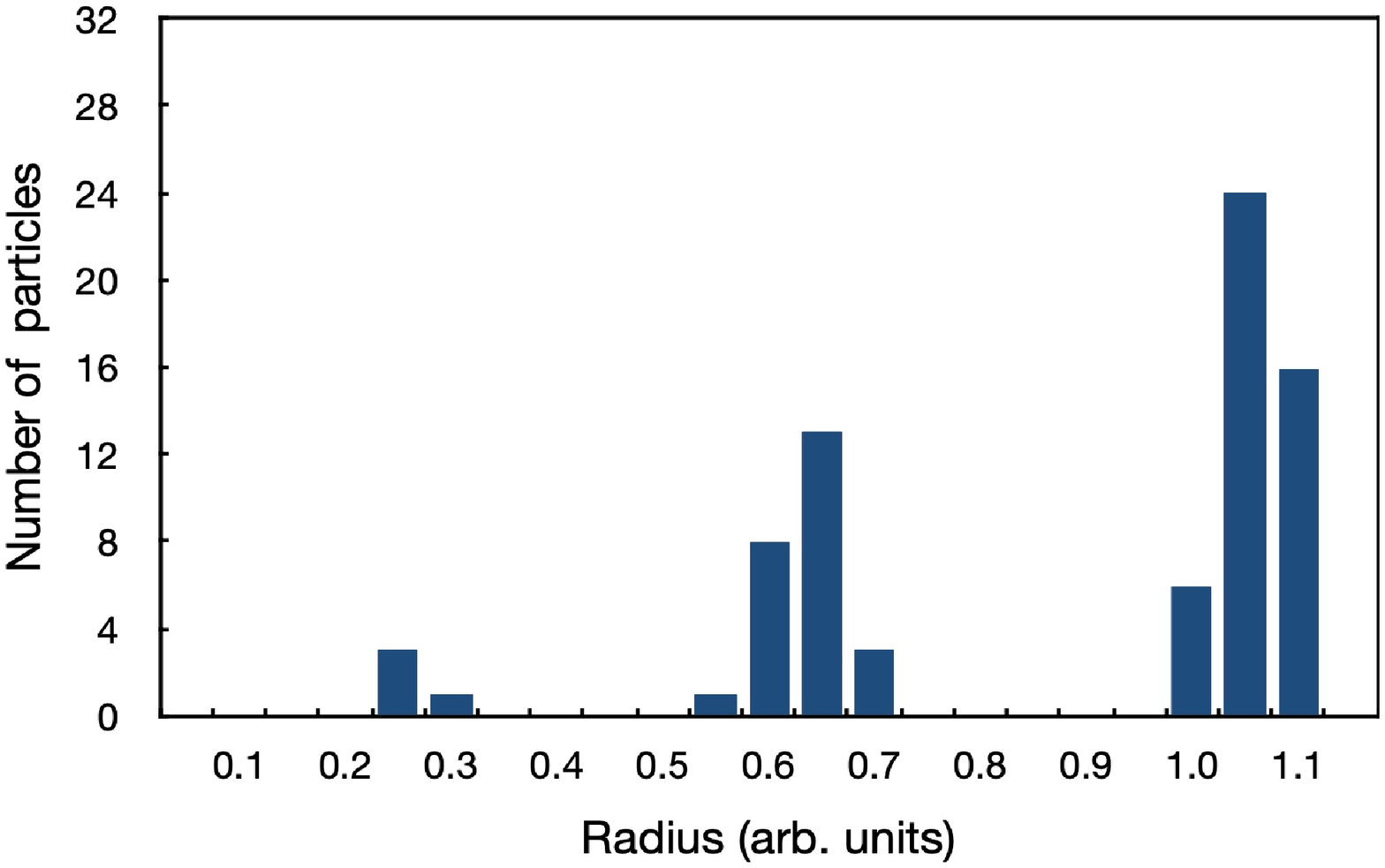}}}
\caption{\label{hist_75} Histogram of the particles' radii for the $A=75$
  system.}
\end{figure}

Figs.~\ref{fig3d_75} and \ref{hist_75} display the optimal configuration
and the histogram of
the particles' radii for the $A=75$ system. The energy is $-538.6275$.
The optimal configuration also shows three distinct shells. The innermost
shell now is formed by a triangular pyramid, which entrenches the spherical
symmetry of the middle and outer shells.

Similarly to the two-dimensional systems, these systems exhibit $SO(3)$ 
symmetry, with the increase in rank indicative of the extra dimension. Angular momentum
emerges once more as a natural symmetry.

\section{Conclusions}
We have generated a catalogue of patterns in two dimensions of systems of
identical particles that optimise
the potential energy, and where the kinetic energy has been neglected.
The systems were allowed to evolve
classically under the action of finite-range attractive potentials with
a repulsive hard core. This allowed for easy
observations of symmetries and correlations. The role played by steric
crowding in such features has been illustrated, with details
of the interaction leading to either crystals or shells.

For the toy 2G potential, the attractive
pocket is smooth and allows deviations from the strict optimal interparticle
distance. Clearly, shells enforce such deviations, whether between
neighbours in a shell or across neighbouring shells at distances not
too far from the strict two particle optimum. Clearly, global binding
then results from interactions between next-to-nearest neighbours  as well
as nearest neighbours.

In the case of the extension to three dimensions, with the use of
the 2G3 potential,
we see the generalisation of the circular symmetry of the optimal
configurations in two dimensions to that of
spherical symmetry in three. The only exception to that is for the two
lightest systems shown, $A=15$ and
$A=20$, where the axis of symmetry and the stronger repulsion of
the 2G3 potential lead to cases of cylindrical symmetry. The increase in
particle number relaxes that symmetry to the spherical symmetry inherent
in the larger systems to allow for the uniform distribution of particles
in each shell.

Given that these are classical systems, evolving dynamically under a
finite-range interaction with short-range attraction and 
a hard repulsive core, one can postulate that the emergence of the shells
stems from an average interparticle distance, as given by the relatively
narrow attraction in the short-range potential. There is no actual 
restriction in the number of particles save for the interparticle distance
aspect eventually saturating one shell and creating 
the next. This is unlike the quantum mechanical case of a system of fermions
with angular momentum symmetry, where 
the number of particles in each shell is determined by the total angular momentum.

Historically, shells were understood in three dimensions from central potential theories,
among which are the exactly solvable Coulomb and harmonic oscillator
potentials. Mean field theories and density functional theories also lead
to shells. Yet in the cases presented in this study, the shells may emerge
without the introduction or use of a centered field approach.
The natural next step is to introduce the kinetic energy, from which we
would also expect the retention of shells, as the action of the potential
would still dictate the interparticle spacing, and also to turn to quantal
systems of identical particles, without and with spin. In the
latter case one may investigate also pairing, and the natural observation
of magic numbers, with angular momentum appearing naturally, given the
spherical, or ellipsoidal symmetry observed herein.

Many cases of shell structures, including the existence of magic numbers, have
been published in the physics of clusters. They are too numerous to be cited
extensively and we refer here to very few of them\cite{Pa93,In86} only. Up to our
knowledge, the majority of cluster studies have used mean field and/or density
functional, centered methods. Our pure two-body approach, with its trivially
simple calculations, may bring new results or easier confirmations and
interpretations of results based on quantum mechanical nuclear structure models. 

In short, the main result of this work is that there is no need of a one-body,
centered theory to bring shells and their symmetries. Bare two-body theories, based on
simple representations of finite-range, local, interactions, are
sufficient and lend a more microscopic basis to the evolution of such
structures and the understanding of interparticle correlations.

\section*{Acknowledgements}
It is a pleasure for B.G.G. to thank B.R. Barrett, 
M. Block, T. Sami, and E. Souli\'e for stimulating discussions.
S.K. acknowledges
support from the National Research Foundation of South Africa.

\bibliographystyle{ws-ijmpe}
\bibliography{vseul}
\end{document}